# Quantifying the quantum nature of high spin YSR excitations in transverse magnetic field

Niels P.E. van Mullekom[1], Benjamin Verlhac[1], Werner M.J. van Weerdenburg[1], Hermann Osterhage[1], Manuel Steinbrecher[1], Katharina J. Franke[2], A.A. Khajetoorians[1]*

**Affiliations:**

[1]*Institute for Molecules and Materials, Radboud University, Nijmegen, The Netherlands*

[2]*Fachbereich Physik, Freie Universität Berlin, Germany*
*Corresponding author: a.khajetoorians@science.ru.nl

Excitations of individual and coupled spins on superconductors provide a platform to study quantum spin impurity models as well as a pathway toward realizing topological quantum computing. Here, we characterize, using ultra-low temperature scanning tunneling microscopy/spectroscopy, the Yu-Shiba-Rusinov (YSR) states of individual manganese phthalocyanine molecules with high spin character on the surface of an ultra-thin lead film in variable transverse magnetic field. We observe two types of YSR excitations, depending on the adsorption geometry of the molecule. Using a zero-bandwidth model, we detail the role of the magnetic anisotropy, spin-spin exchange, and Kondo exchange. We illustrate that one molecular type can be treated as an individual spin akin to an isolated spin on the metal center, whereas the other molecular type invokes a coupled spin system represented by a spin on the center and the ligand. Using the field-dependent evolution of the YSR excitations and comparisons to modeling, we describe the quantum phase of each of the molecules. These results provide an insight into the quantum nature of YSR excitations in magnetic field, and a platform to study spin impurity models on superconductors in magnetic field.

An individual spin impurity exchange coupled with a superconductor can lead to local in-gap excitations referred to as Yu-Shiba-Rusinov (YSR) excitations [1-3]. These in-gap excitations define the energy difference between binding or unbinding a quasiparticle to the spin impurity. The excitation energy depends primarily on the competition between the superconducting pairing energy ($\Delta$) and the Kondo exchange energy ($J_K$) and defines two distinct regimes that refer to either the binding or unbinding of a quasiparticle. The transition between these two regimes is often referred to as the quantum phase transition (QPT). Experiments based on scanning tunneling microscopy/spectroscopy (STM/STS) have been widely successful at studying YSR excitations derived from atomic and molecular impurities[4-6]. Nevertheless, most experiments were performed in the absence of a magnetic field. Yet, methods that involve magnetic field are essential to determine the parity of the quantum states involved in the excitation, namely the "excitation pathways," as well as the spin state of the impurity. In contrast to high-spin impurities, for $S = 1/2$, the excitation pathway as a function of magnetic field simplifies[7,8]. This is due multiple interactions, such as interatomic exchange, competing Kondo exchange energies and single-ion magnetic anisotropy can be neglected. Yet a vast majority of spin impurities on surfaces are derived from $3d$ or $4f$ atomic spins[9-11], where the total spin $S > 1/2$, necessitates consideration of competing energy scales on the excitation pathways.

Going beyond the $S = 1/2$ picture necessitates a quantum description that considers multiple energy scales, such as the Kondo exchange coupling in a number of channels (i.e. $J_{K_i}$), intra-atomic exchange (e.g. Hund), on-site Coulomb energies (e.g. $U$), and single-ion magnetic anisotropy[12]. Spin impurity models are the most prevalent way to quantify the role of various interactions on the resultant excitation pathways[13-15]. Most often, the description is reduced to a giant spin model, neglecting Hund's exchange and the interplay of Coulomb interactions, and this giant spin is coupled to a bath. On a superconductor, the latter is often treated in a zero-bandwidth model which shows that the role of renormalization[16] is rather weak, thus further reducing the computational complexity of the problem. Based on these model predictions, the excitation pathways of high-spin impurities is determined by a sensitive interplay between multiple energy scales, which can only be discriminated by systematically modifying an energy scale, for instance by a magnetic field[13,14]. Certain experimental approaches have been used to modify the Kondo exchange energy in a limited range, enabling identification of the excitation pathways[17,18]. While an applied magnetic field would be an ideal perturbation[7,8], the upper critical field of typical BCS superconductors used in experiments corresponds to an energy much smaller than the desired Zeeman energies needed to observe changes in the YSR excitations.

Here, we quantify the response of YSR excitations of high-spin impurities to a large transverse magnetic field. We start by depositing individual manganese phthalocyanine (MnPc) molecules on the surface of a quantum-confined and superconducting lead (Pb) film. MnPc molecules exhibit two distinct types of YSR excitations, depending on the binding site and orientation of the molecule with respect to the substrate. Due to the combination of quantum confinement and spin-orbit coupling, we observe negligible changes to the superconducting gap structure in response to magnetic fields parallel to the surface, up to $B_\parallel \leq 4$ T. Using the robustness of the superconductor, we quantify the changes of all the YSR excitations for each molecule type in the presence of an applied transverse magnetic field up to $B_\parallel = 4$ T. Unlike the expected behavior for a $S = 1/2$ impurity, we observe a non-linear and non-monotonic evolution of the YSR excitations for both molecule types. We additionally observe multiple YSR excitations, and a change of the total number of these excitations in magnetic field. Using a zero-bandwidth spin model, which considers $J_K^i$, single-ion anisotropy, magnetic exchange, and the Zeeman energy, we quantify the role of these various interactions on the YSR excitations and the excitation pathways. Based on this, we identify trends in the model simulations, which reproduce parts of the observed spectra. We also illustrate that the conventional model descriptions fail to capture vital aspects observed in the experiment in an applied magnetic field, suggesting that new theoretical understandings that consider transport-based effects and go beyond the zero-bandwidth picture may be necessary in order to understand the YSR problem in magnetic field.

In order to study YSR excitations in magnetic field, we started by depositing ultra-thin Pb films on the surface reconstruction Si(111)-Ag ($\sqrt{3} \times \sqrt{3}$). A schematic of the experiment is illustrated in Fig. 1a (see methods and SI1). We intentionally worked with ultra-thin Pb films due to their extraordinarily large in-plane upper critical field ($H_c^2$)[19,20]. We observed a robust hard superconducting gap, which was insensitive to the value of $B_\parallel$, up to 4 T (Fig. S2). Using these Pb films, we subsequently deposited MnPc onto the surface. Individual MnPc molecules preferentially absorb on the step edges of the film. By using lateral manipulation with the STM tip (see methods), we dragged individual molecules onto chosen locations on a given terrace of the grown Pb film. All the subsequent spectra are measured on the center of the molecule. We observed two types of YSR spectra, at zero magnetic field, as displayed in Fig. 1c,d. These spectra depend primarily on the given binding site and orientation of the molecule relative to the underlying atomic lattice of Pb(111). We refer to the two cases as, MnPc1, where one of the ligand axes of the molecule is parallel to one of the high symmetry directions of the Pb(111) film, and MnPc2, where the ligand axes of the molecule are bisected by one of the high symmetry directions of the Pb(111) film (Fig. S3). The YSR excitations of MnPc1 feature one pair of peaks with larger intensity at positive bias voltage, whereas MnPc2 shows three pairs of peaks with larger intensity at negative bias. The inversion in the asymmetry of the YSR excitations between the two types is typically considered attributed to inverting the excitation pathway. On bulk Pb(111), both isolated and MnPc molecules in a densely packed monolayer illustrate a three-peak structure[21]. In that case, the three YSR excitations were ascribed to the excitation of anisotropy-split states from $S = 1$ molecules. The energy and the intensity of the states differed depending on adsorption site. This variation was attributed to changes in the excitation pathway of the molecular spin, due to a variation in $J_K$. Likewise, Kondo-like resonances have been observed for MnPc on both Pb(111) bulk as well as Pb thin films [6,22], which qualitatively agree with the Kondo-like spectra taken on MnPc1 and MnPc2 (Fig. S4).

Before exploring the observed magnetic field-dependence of both MnPc1 and MnPc2, we describe the expected experimental behavior based on theoretical modeling[14,23]. We employ a zero-bandwidth model with both one (two) superconducting site(s) and one (two) spin site(s), as previously described in refs. [14,23]. Such modeling can capture the interplay between the YSR excitations, including various energy scales, such as exchange couplings and magnetic anisotropy. We additionally consider a Zeeman term and a transverse anisotropy (see methods), and neglect particle-hole asymmetry in the simulation. An example of this is illustrated in Fig. 2b, in which the excitation diagram as a function of the magnetic field can be linked to the measurements. It has been previously shown that such models capture the qualitative physics, as renormalization effects are typically weak in the YSR problem [14,16]. Our choice of considering two spin sites with magnetic anisotropy was motivated by previous *ab initio* electronic structure calculations of an individual MnPc molecule on Pb(111)[24,25], aimed at understanding the Kondo behavior in this system[6,22]. Based on this input, it was shown that the Mn atom hosts a total spin $S > 1/2$, due to Hund's exchange, and that there is a significant crystal field splitting. Moreover, in ref. [25], it was further shown that the ligands can acquire a spin, that antiferromagnetically couples to the Mn atom.

Based on these two models and the exploration of their parameter space, we can distinguish three categories of magnetic field-dependent trends: (1) A field-dependent splitting of the YSR excitations that are degenerate at $B = 0$ T. This occurs when the excited state is Kramer's degenerate, or when the excited state has integer spin without magnetic anisotropy leading to a doubly degenerate state (e.g. $|\pm 1\rangle$) that can be accessed via selection rules. (2) A non-linear $B$-dependent evolution of a given YSR excitation. This occurs when rotational symmetry is broken, due to either magnetic anisotropy for $S > 1/2$, or when there is a difference in $g$-factor between two coupled spins. (3) A change in the number of YSR excitations for $B_\parallel \neq 0$ T. A change in the total number of YSR excitation is usually associated with a QPT of the ground state. For $S > 1/2$, there are two types of QPTs to consider. (a) A change in the fermion parity of the ground state, namely a change in the number of bound quasiparticles. This change in parity is always accompanied by a YSR excitation that crosses the gap center. (b) A parity preserving QPT that changes the ground state spin projection of the total ground state, which may lead to a change in number of accessible excitations due to selection rules [14] (Fig. S5).

Next, we review the measured YSR states of individual MnPc1 molecules as a function of $B_\parallel$. In Fig. 3a, we plot the spectra in a high-resolution false color plot with increasing transverse field strength. For the subsequent descriptions, we focus on the subset of YSR excitations at one bias polarity to avoid confusion. For $B_\parallel < 0.5$ T, the sole YSR excitation shows an overall insensitive response to applied magnetic field. Around $B_\parallel \approx 0.5$ T, the YSR excitation splits asymmetrically, with an asymmetric spectral weight favoring the state closer to the gap center. For $B_\parallel > 0.5$ T, the excitation with higher intensity shows a non-linear evolution first trending toward the gap center, then after an inflection point moving toward the gap edge. The other split excitation at higher energy shows an almost linear evolution with a different slope from the former state and loses intensity as it approaches the gap edge. We note that none of the excitations crosses the gap center.

The $B_\parallel$-evolution of the YSR states is a signature of the interplay of a transversal magnetic field and magnetic anisotropy. In Fig. 3b, we illustrate the modeled YSR excitations for a single spin $S = 1$ with axial and transverse anisotropy, coupled to a single SC site in a transverse magnetic field. We note that the modeled YSR excitations are very sensitive to the interplay of the various parameters, and we chose the model parameters that best reproduce the experimental spectra (see Fig. S6). We also display the corresponding energy-level diagram (Fig. 3c). The scenario corresponds to a bound quasiparticle in the ground state, namely a (partially) screened ground state. At $B = 0$, the models yield one YSR excitation that corresponds to the transition from the ground state to the Kramer's degenerate doublet. Due to magnetic anisotropy, the ground state energy is insensitive to $B_\parallel < 0.5$ T. However, the excited state is a spin-½ doublet that is affected by $B_\parallel$ resulting in a splitting and the observed two branches. This observed splitting occurs at a finite value of $B_{\parallel}$ due to the considered broadening in the calculation (see methods). As $B_{\parallel}$ increases, the Zeeman energy becomes larger than the magnetic anisotropy and the ground state aligns with $B_{\parallel}$. This results in a non-linearity in the $B_{\parallel}$-dependence of the ground state and produces an inflection point. Although the simulation captures the observed trends well, there are other parameter sets that result in a qualitatively similar trend (Fig. S6). However, in all the cases we simulated, the best match of parameters always yields a ground state with the same parity, i.e. no bound quasiparticles.

Next, we discuss the $B_{\parallel}$-evolution of the YSR excitations of MnPc2. We characterized two distinct cases MnPc2($\alpha = 15°$) and MnPc2($\alpha = 45°$), where $\alpha$ corresponds to the smallest angle of a molecular ligand axis and $\alpha = 0$ corresponds to $B_{\parallel}$ being parallel to a ligand axis (white dashed line in the inset of Fig. 1d) with respect to $B_{\parallel}$ (red arrow). We note that the orientation of $B_{\parallel}$ remains fixed. In Fig. 4a, we illustrate the typical case for MnPc2($\alpha = 15°$). With increasing $B_{\parallel}$ we observe a peak that hardly shift except for a small bowing with an inflection point toward the gap edge $B_{\parallel} \approx 2$ T. It also increases both in its intensity and linewidth for $B_{\parallel} > 0.5$ T, where a peak splitting off from the second peak (ii) merges in. Not only that peak (ii) splits at about $B_{\parallel} \approx 0.5$ T, but also the outermost peak (iii) is clearly split at this applied magnetic field (red arrows). At the intermediate field of $B_{\parallel} \approx 0.5$ T, we thus detect in total five YSR excitations. From the different splitting of both peaks (ii) and (iii), we conclude that the excitations must involve states with different spin projections. Above $B_{\parallel} \approx 1$ T, the inner branch of peak (ii) is not discernable anymore from resonance (i), while also the outer branch from (ii) and the inner one from (iii) merge (magenta arrows), reducing the number of observable YSR states to three. The other respective branches of peaks (ii) and (iii) move toward the gap edge with different slopes. The outer branch of peak (iii) crosses the coherence peak at $B_\parallel \approx 2.5$ T (yellow dashed line) and continues outside the gap (yellow arrow). We note that when the branch of peak (iii) crosses the coherence peak, there are no observed sudden changes in the YSR spectra. Finally, we do not observe a crossing or splitting of the merged peaks up to $B_\parallel \approx 4$ T (Fig. S7). This observation is contrary to the expectation of YSR excitations originating from different spin states that cross. It is also surprising that the split-off branch from peak (ii) does not seem to continue after it merged with (iii).

In Fig. 4b, we illustrate the typical case for MnPc2($\alpha = 45°$). Peaks (i),(ii) illustrate a nearly linear shift toward the gap edge with slightly different slopes, until $B_\parallel \approx 2.5$ T. The outer peak (iii) remains relatively insensitive to a magnetic field up to $B_\parallel \approx 0.5$ T and then subsequently splits (red arrow). The branch of peak (iii) that moves toward the gap center, merges with the shifted peaks (i) and (ii) at $B_\parallel \approx 2.5$ T (magenta arrow). As also seen for MnPc2($\alpha = 15°$), the other branch of peak (iii) crosses the location of the coherence peak (yellow dashed line) and persists outside of the gap in increasing values of $B_\parallel$, with diminishing intensity. The crossing of the branch of peak (iii) occurs at lower values of $B_\parallel$, compared to the merging of all the other observed states (magenta arrow). The merging of these peaks remains up to $B_\parallel = 4$ T, (Fig. S7), leading to a broadening of the overall linewidth and intensity. This is like the previous case, and this merging persists for $\Delta B \approx 1.5$ T without the observation of a crossing or splitting of states. We note that the intensity of the branch of state (iii) observed outside of the gap decreases.

The observation that both orientations of MnPc2 show distinctly different $B_\parallel$-dependence illustrates the importance of transversal anisotropy and thus their quantum nature of the YSR excitations. For both orientations of MnPc2, we modeled the YSR excitations as two spin sites with $S_1 = 1$, representing the spin on the Mn center and $S_2 = 1/2$, representing the spin on the ligand. As the molecules lie in different orientations with respect to the underlying Pb lattice, the electronic structure of MnPc molecule, including its crystal field and hybridization may vary. This can be mimicked by considering different values of $D, E, J_K, S$, compared to MnPc1. In Fig. 4c,d, we illustrate the modeled YSR spectra for MnPc2($\alpha = 15°$) and MnPc2($\alpha = 45°$), respectively, where we considered a spin $S_1 = 1$ with magnetic anisotropy coupled antiferromagnetically with a second spin $S_2 = 1/2$ in a transverse magnetic field. We also display the corresponding energy-level diagrams (Fig. S8). Due to the impact of magnetic anisotropy on $S_1$, the simulation yields three YSR excitations for $B_\parallel = 0$ T. In this case, the excitations labeled (ii) and (iii) are $2E_1$ and $D_1 + E_1$ higher in energy than (i), respectively, yielding three YSR states. $S_2$ creates two YSR excitations that are Kramer's degenerate, as well as nearly degenerate with the highest excitation of $S_1$ at $B_\parallel = 0$ T, yielding a total of three peaks at $B_\parallel = 0$ T.

Next, we consider $B_\parallel > 0$ T, for MnPc2($\phi_B = 60°$) in Fig. 4c. Excitation (i) first increases in energy and then becomes nearly constant for increasing $B_\parallel$. Excitation (ii) follows the same trend as (i) with a slightly different slope. Peak (iii), which originates from both the third excitation of $S_1$ and the Kramer's pair from $S_2$, splits linearly in energy as $B_\parallel$ increases (red arrow) and crosses state (ii) at $B_\parallel \approx 3.5$ T (magenta arrow). The number of observable YSR states thus changes from three ($B_\parallel < 0.5$ T) to four (0.5 T $< B_\parallel < 3$ T) and finally to three ($B_\parallel > 3$ T).

Finally, we consider $B_\parallel > 0$ T, for MnPc2($\phi_B = 90°$) in Fig. 3d. Excitation (i) increases linearly in energy for increasing $B_\parallel$, in contrast to the $\phi_B = 60°$ case. Excitation (ii) increases in energy and has an inflection ($B_\parallel \approx 1$ T), resulting in a crossing of the peaks (i) and (ii) at $B_\parallel \approx 3$ T (magenta arrow). To reproduce the one broad peak inside of the gap, we intentionally chose $g_1 < g_2$ such that this crossing occurs at a higher value of $B_\parallel$ and in tandem with the overlap of the YSR excitation stemming from the $S_2$ (magenta arrow). Peak (iii) evolves the same as for MnPc2($\phi = 60°$). The number of observable YSR states thus changes from three ($B_\parallel < 0.5$ T) to four (0.5 T $< B_\parallel < 3$ T) and finally to two ($B_\parallel > 3$ T).

While the model simulations of MnPc1 agree reasonably well with the experimental data, there are also clear differences in the simulated spectra and the experimental spectra with respect to MnPc2. There are three classes of observations that cannot be reproduced in the model for MnPc2: (a) certain non-linear splitting, (b) the robustness of merged states in variable magnetic field, and (c) the physical nature of the observed out-of-gap excitations. As for (a), the model predicts a splitting of excitation (iii) that is symmetric in energy (red arrow), while in the experiment for MnPc2($\alpha = 45°$) we observe a non-linear evolution of the split peaks, each of which has a different slope. In addition, the simulation does not capture the observed splitting of peak (ii) for MnPc2($\alpha = 15°$) around $B_\parallel \approx 0.5$ T (red arrow). Finally, neither the intensities of the YSR excitations at $B = 0$ T nor the evolution of the intensities are captured correctly in the simulation. This may be due to state-dependent and potentially magnetic-field-dependent scattering as well as co-tunneling through molecular orbitals[26], which are not considered.

Concerning (b), based on the model parameters, we cannot reproduce a merging of YSR excitations that remain merged for significant changes in $B_\parallel$ (magenta arrows). This is most striking for $\alpha = 45°$, where around $B_\parallel \approx 2.5$ T, these merges remain robust up to $B_\parallel = 4$ T (Fig. S7). This merging suggests that the expected excitations cannot be treated independently once the states become degenerate. Finally, (c), the model will not properly capture the continuation of the YSR excitations near the gap edge into the quasiparticle continuum (yellow arrows), and its effect on the YSR excitations that remain inside the gap. We further discuss these points, below. However, these differences cannot be reproduced by considering substantially different values of the various model parameters, considering the constraint that both orientations of MnPc2 should stem from the same set of parameters.

The inability to capture the mentioned differences in the modeling of YSR excitations, illustrates both the quantum nature of the problem, and the potential need to consider typically neglected effects. As described above, the robust merging of the YSR excitations in variable magnetic field for both orientations cannot be completely described by exchange-coupled two-site excitations within the zero-bandwidth model (magenta arrows in Fig. 4a,b). Due to the nature of the YSR excitation, which links spin states of different spin projection, the Zeeman effect will lead to states crossing rather than merging. This can change in the case of a QPT, or a spin transition of a given multiplet. An abrupt change in YSR states can occur near a QPT, which can result from either a change in fermion parity or a high/low spin transition in the ground state. However, a QPT in fermion parity should be accompanied by one of the YSR excitations crossing the gap center which we do not observe. It has been observed that higher order tunneling processes can lead to additional states near such a QPT (e.g. for $S = 1/2$)[8]. However, it is not clear why higher order processes would lead to a merging of states that persist for the observed ranges of $B_\parallel$. Additionally, we did not consider any non-equilibrium effects, like spin pumping. In the case of pumping, simplistically, additional states would be expected to emerge, similar to higher order tunneling processes, and be dependent on, for example, $I_t$. Moreover, the nature of the QPT is not well-defined when spin rotation symmetry is broken, due to transverse anisotropy or a transverse magnetic field. This is a result of the quantum nature of the problem, in which there is not a well-defined quantum number that describes the relevant spin states.

Finally, the nature of the excitations that extend outside of the superconducting gap and their impact on the coincident YSR excitations that remain in the gap are also not known. Traditionally, these can be explained by inelastic excitations[27-29]. Pair excitations out of the gap have been observed at zero field[30]. However, in this case, it is unclear what the role of spin-orbit coupling combined with the applied magnetic field are on such excitations, as SU2 symmetry is broken. These observations suggest that it may be important in future theoretical modeling to consider effects like renormalization, bath hopping, multi-electron processes, and co-tunneling in these types of experiments.

In conclusion, we track the evolution of YSR excitations between quantum states of a high-spin molecule using transversal magnetic field. We observe two different types of YSR excitations, which are dictated by the orientation of the molecular ligands with the underlying Pb(111) substrate. In the case of MnPc1, we demonstrate that the YSR excitations can be modeled by one individual high spin with magnetic anisotropy. This suggests that the relevant spins stem from the Mn atom. In contrast, for the case of MnPc2, which is a different orientation of the molecule with respect to MnPc1, the YSR excitations can only be modeled considering multiple coupled spins, similar to previous reports[31]. This suggests that in this orientation, the spin polarization produced on the ligands cannot be neglected. Based on the combination of experimental data and theoretical modeling, we are able to qualitatively reproduce many of the observed features, and to relate this to the excitation pathway of ground state and the various excited states of the molecular type, as well as the relevant magnetic properties of the molecule. Nevertheless, this modeling, as seen for the case of MnPc1, does not necessarily lead to a unique set of parameters of spin and magnetic anisotropy. In the case of MnPc2, the model is not able to reproduce all the observed features, regardless of the parameter space we have explored. These observations motivate considering modeling that goes beyond, for example, the effects of co-tunneling[32] on the atomic spin excitations, the presence of unexpected excitations and non-trivial electron-electron interactions, as well as relevant Kondo renormalization and spin pumping effects. These results present a pathway to study spin impurity models on superconductors in magnetic field, and to explore the quantum nature of these excitations and their dynamics.

**Methods**

We used a home-built UHV STM/STS facility, which operates at $T = 30$ mK, where a magnetic field can be applied either perpendicular as well as parallel to the sample surface [33,34]. STS was taken using a lock-in method, with the modulation voltage applied to the sample. Si(111)−(7 × 7) was prepared by repetitive flashing to $T \sim 1450°C$, as measured by a pyrometer aimed at the sample surface. Ag was subsequently deposited at room temperature (RT) and post-annealed at $T \sim 575$ °C for 15 minutes to form the reconstruction Si(111)-

Ag($\sqrt{3} \times \sqrt{3}$) (see Fig. S1). Next, Pb was deposited while the substrate was cooled on a liquid nitrogen cold stage ($T \sim 110$ K). Finally, the MnPc molecules were deposited on the substrate at RT, after which the substrate was transferred into a He flow-cryostat cooled transfer arm to stop the RT anneal time (4 minutes from cold stage to transfer arm) and transferred into the STM.

The prominent moiré structure[35] and the resultant quantum well states (QWS)[36,37] of the Pb films are detailed in Fig. S1. We focused on Pb films that were 11 ML thick. At smaller energy scales, we observed a superconducting gap with an extracted gap value of $\Delta = 1.29 \pm 0.01$ mV using a normal-metal tip, similar to previous reports using STS[5,38]. The use of a normal tip combined with the measurement temperatures allows to eliminate the use of a superconducting tip, and the necessary field-dependent deconvolution of its states on the spectra. In Fig. S2 we illustrate a typical spectrum of the Pb surface as a function of in-plane magnetic field ($B_{\parallel}$) up to $B_{\parallel} = 4$ T, measured > 4 nm away from any MnPc molecule center. MnPc molecules were laterally manipulated from the step edge of the Pb films, and intentionally placed at various locations along the film surface. Lateral manipulation was performed with the following parameters (feedback closed, $V_s = 100 - 200$ mV, $I_s = 1 - 2$ nA).

The zero-bandwidth model used in this paper is composed terms for the substrate, YSR interaction, magnetic anisotropy and exchange interaction all of which are described in detail by ref. [14,23,30] as well as additional terms for the Zeeman energy of the spins sites and superconducting site:

$$H = H_{\text{SC}} + H_{J_\text{K}} + H_{\text{MAE}} + H_{J_{\text{ex}}} + H_{\text{Zee}}.$$

For the two spin and two superconducting sites model this is:

$$H_{\text{SC}} = \sum_{i=1}^{2} \Delta \left( c_{i,\uparrow}^{\dagger} c_{i,\downarrow}^{\dagger} + c_{i,\downarrow} c_{i,\uparrow} \right) - \sum_{\sigma=\uparrow,\downarrow} t \left( c_{1,\sigma}^{\dagger} c_{2,\sigma} + c_{2,\sigma}^{\dagger} c_{1,\sigma} \right),$$

$$H_{J_{\text{K}}} = \sum_{i=1}^{2} \sum_{\sigma=\uparrow,\downarrow} \sum_{\sigma'=\uparrow,\downarrow} c_{i,\sigma}^{\dagger} (\mathbf{S}_i \cdot \hat{J}_{\text{K}_i} \cdot \mathbf{s}_{\sigma,\sigma'}) c_{i,\sigma'},$$

$$H_{\text{MAE}} = \sum_{i=1}^{2} D_i S_{i,z}^2 + E_i (S_{i,x}^2 - S_{i,y}^2),$$

$$H_{J_{\text{ex}}} = \mathbf{S}_1 \cdot \hat{J}_{\text{ex}} \cdot \mathbf{S}_2,$$

$$H_{\text{Zee}} = -\sum_{i=1}^{2} g_i \mu_{\text{B}} \boldsymbol{B} \cdot \boldsymbol{S}_i - \sum_{i=1}^{2} \sum_{\sigma=\uparrow,\downarrow} \sum_{\sigma'=\uparrow,\downarrow} c_{i,\sigma}^{\dagger} (g_{\text{SC}} \mu_{\text{B}} \boldsymbol{B} \cdot \mathbf{s}_{\sigma\sigma'}) c_{i,\sigma'}.$$

Here $\boldsymbol{S}_i$ is the vector of spin operators for spin site $i$, and $\boldsymbol{s} = \frac{1}{2} \boldsymbol{\tau}$ in terms of the vector of Pauli matrices $\boldsymbol{\tau}$. We note that without the Zeeman field on the superconducting sites, the $B$-field evolution is non-linear[13]. For the one spin, one superconducting site, $i = 1$, $t = 0$ and $H_{J_{\text{ex}}} = 0$.

To obtain the simulated YSR spectra, for each value of $\boldsymbol{B}$, we calculate the transition coefficient for every excited state:

$$a = \sum_{i=1}^{2} \sum_{\sigma=\uparrow,\downarrow} \sum_{\lambda} \left| \langle \lambda | c_{i,\sigma}^{\dagger} | \text{gs} \rangle \right|^2 + \left| \langle \lambda | c_{i,\sigma} | \text{gs} \rangle \right|^2,$$

where $|\lambda\rangle$ and $|\text{gs}\rangle$ denote the excited and ground state respectively and subsequently plot two gaussians at $\pm(E_\lambda - E_{\text{gs}})$ with amplitude $a$ and FWHM = 0.1 mV. Finally, all gaussian peaks are summed resulting in the simulated spectrum $A(E, \boldsymbol{B})$.

**Acknowledgements**

We would like to thank Christian Ast, Malte Rösner, Daniel Wegner, Jacob Linder, and Mikhail Katsnelson for fruitful discussions:

**Author contributions**

N.P.E.vM., B.V., W.M.J.vW, H.O., and M.S. performed the experiments. N.P.E.vM. and B.V. developed the model calculations with N.P.E.vM. carrying out the calculations in the manuscript. K.J.F. and A.A.K. designed the experiments, and A.A.K. additionally participated in the experiments. N.P.E.vM., B.V., W.M.J.vW, H.O., and M.S. all performed the experimental analysis, with all authors participating in the discussion of the results of the analysis and its further iterations. N.P.E.vM and A.A.K. primarily wrote the manuscript, while all authors provided input during its development.

**Funding**

This project was supported by the European Research Council (ERC) under the European Union's Horizon 2020 research and innovation program (grant no. 818399). B.V. acknowledges funding from the Radboud Excellence fellowship from Radboud University in Nijmegen, the Netherlands. This publication is part of the project "What can we 'learn' with atoms?" (with project number VI.C.212.007) of the research program VICI which is (partly) financed by the Dutch Research Council (NWO). This publication is part of the project TOPCORE (project no.

OCENW.GROOT.2019.048) of the research program Open Competition ENW Groot, which is partly financed by the Dutch Research Council (NWO). K.J.F. also acknowledges support from the Deutsche Forschungsgemeinschaft through grant FR2726/10-1.

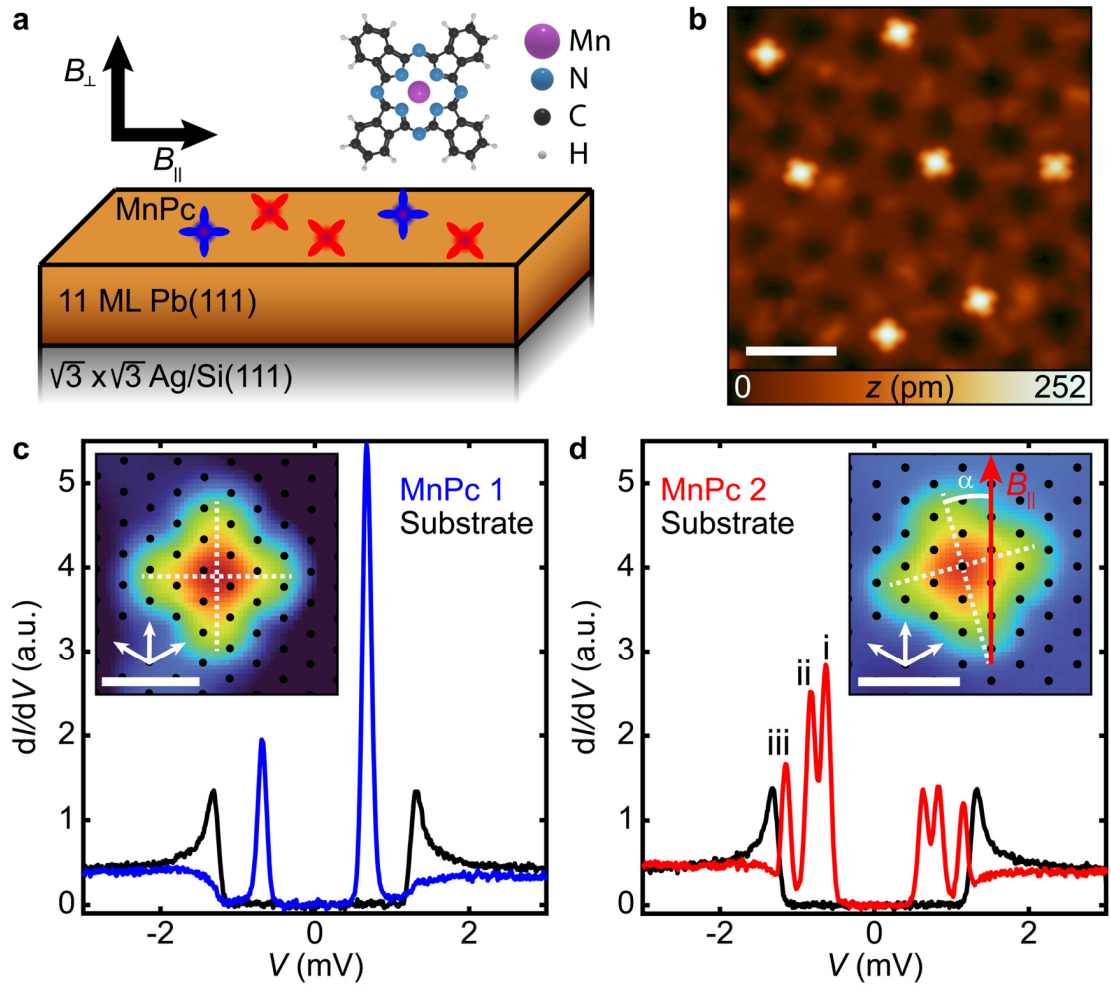

**Figure 1: YSR excitations of individual MnPc molecules on the surface of an 11 ML Pb film:** (a) schematic of the sample system studied, composed of individual MnPc molecules, an 11 ML Pb film grown on a Si(111)-Ag ($\sqrt{3} \times \sqrt{3}$) surface. (b) Constant-current STM image of MnPc deposited on 11 ML of Pb ($V_S$=90 mV, $I_t$=5 pA; scale bar, 5 nm). (c) YSR excitations at $B$ = 0 T of a typical MnPc1 type molecule (blue). The substrate spectrum (black) was measured 4 nm away from the molecule. MnPc1 is characterized by having one of its ligand axes parallel to one of the high symmetry directions of the Pb(111) film (inset: MnPc superimposed over the Pb(111) lattice (black) and the white dashed lines indicate ligand axes, white arrows indicate high symmetry directions (scale bar, 1 nm)). (d) YSR excitations at $B$ = 0 T of a typical MnPc2 type molecule (red). The substrate spectrum (black) was measured 11 nm away from the molecule. MnPc2 is characterized by having one of its ligand axes bisected by one of the high symmetry directions of the Pb(111) film (inset, similar to (c), the field angle ($\alpha$) is indicated with respect to the ligand axes (scale bar, 1 nm)). All spectra were measured with $V_S$ = 6 mV, $I_t$ = 200 pA, $V_{\text{mod}}$ = 20 μV, $T$ = 30 mK and a W tip.

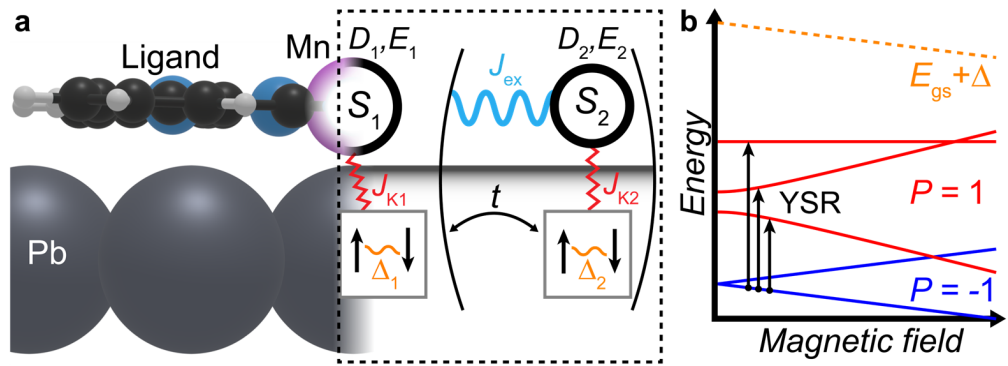

**Figure 2: Zero-bandwidth model of the YSR excitations in magnetic field**
(a) schematic side view of the MnPc molecule that links the physical system to the spin/superconducting sites in the model (see methods). The model considers spin values ($S_i$), Kondo exchange ($J_{K_i}$), spin-spin exchange coupling ($J_{ex}$), magnetic anisotropy ($D_i, E_i$), and hopping between superconducting sites ($t$). (b) Example of an energy level diagram for the one site model with $S = 1$, with the ground state (bound quasiparticle, parity $P = -1$) and the excited states (unbound quasiparticle, $P = 1$), easy-axis and transverse anisotropy in a magnetic field, illustrating how this level structure relates to the measured YSR excitations.

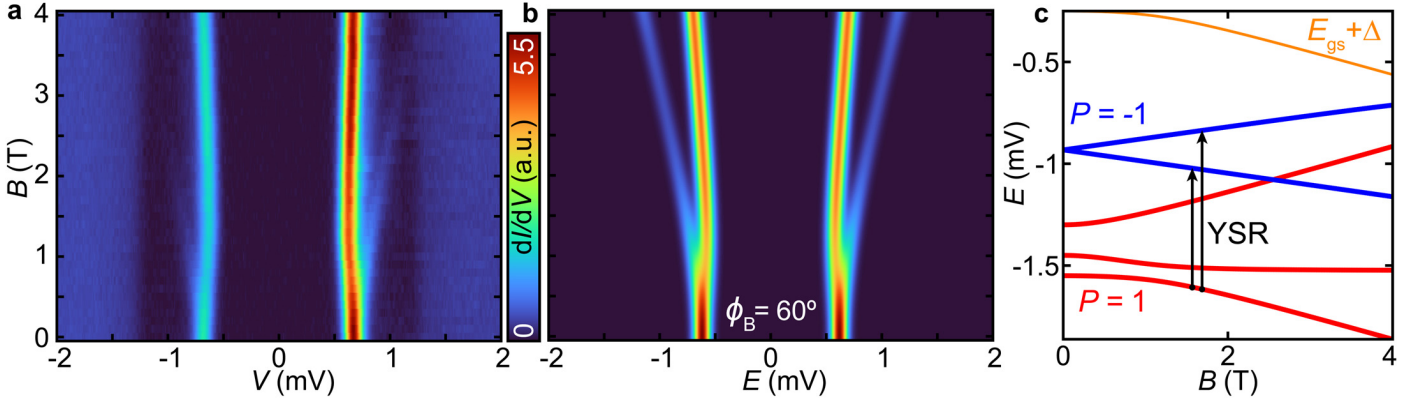

**Figure 3: Transverse magnetic-field dependence of the YSR excitations of MnPc1 given by a single spin and a single site:** (a) a false-color plot of the STS spectra taken at $B_\parallel$ in steps of $\Delta B_\parallel = 0.1\ T$, up to $B_\parallel = 4\ T$. All data was measured with $V_S = 6$ mV, $I_t = 200$ pA, $V_{\text{mod}} = 20\ \mu$V, $T = 30$ mK and a W tip. (b) Simulation of the YSR spectra using the one site model with parameters (see methods for Hamiltonian): $S = 1$, $g = 2$, $D = -0.2$ mV, $E = -0.05$ mV, $\Delta = 1.3$ mV, $g_{SC} = 2$, $J_K = 0.79$ mV, and transverse magnetic field defined using a polar angle $\theta_B = 90°$ (from $+\hat{z}$ to $\hat{B}$), and azimuthal angle $\phi_B = 60°$ (from $+\hat{x}$ to the orthogonal projection of $\hat{B}$ on the $x$-$y$-plane). (c) associated Zeeman level diagram of simulated YSR spectra in (b). Energy eigenstates in red (blue) belong to parity $P = 1$ (bound) ($P = -1$ (unbound)). Orange line indicates the energy of the ground state $E_{gs} + \Delta$.

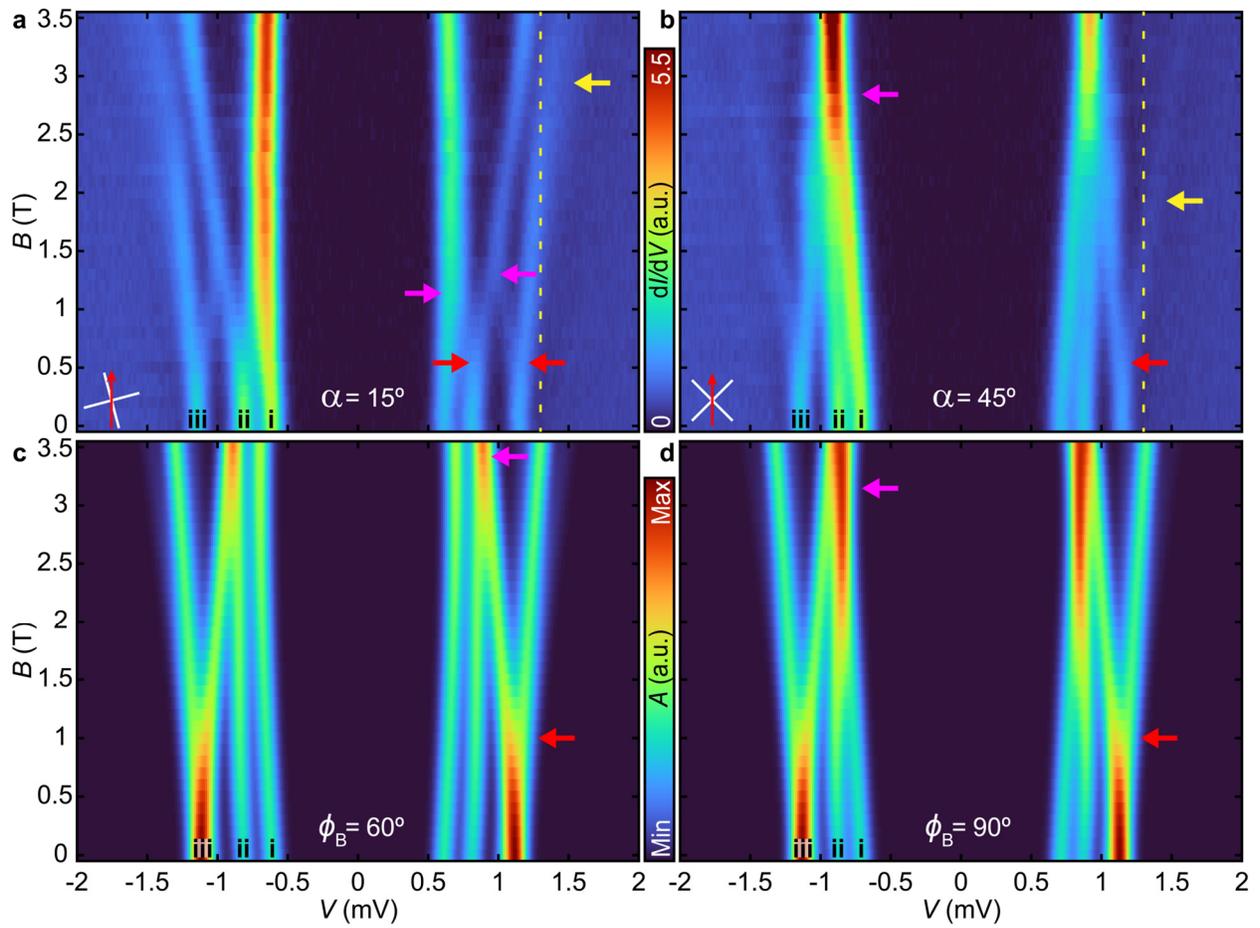

**Figure 4: Transverse magnetic field dependence of the YSR excitations of MnPc2 given by coupled spins and two sites.** (a) a false-color plot of the STS spectra of MnPc2($\alpha = 15°$) taken at $B_\parallel$ in steps of $\Delta B_\parallel = 0.1$ T , up to $B_\parallel = 3.5$ T ($V_S = 6$ mV, $I_t = 200$ pA, $V_{mod} = 20$ $\mu$V), with $\alpha$ defined as in Fig. 1d. (b) a false-color plot of the STS spectra of MnPc2($\alpha = 45°$) taken at $B_\parallel$ steps of $\Delta B_\parallel = 0.1$ T, up to $B_\parallel = 3.5$ T. (c) Simulation of the YSR spectra using the two site model with parameters: $S_1 = 1$, $S_2 = 1/2$, $g_1 = 1.5$, $g_2 = 2$, $D_1 = -0.43$, $E_1 = -0.1$, $D_2 = E_2 = 0$, $J_{ex} = 0.05$, $\Delta_{1,2} = 1.3$, $g_{SC} = 2$, $t = 0$, $J_{K_1} = 2.14$, $J_{K_2} = 3.20$, $\theta_B = 90°$ and $\phi_B = 60°$. (d) Same as (c) but with parameters: $S_1 = 1$, $S_2 = 1/2$, $g_1 = 1.5$, $g_2 = 2$, $D_1 = -0.35$, $E_1 = -0.08$, $D_2 = E_2 = 0$, $J_{ex} = 0.05$, $\Delta_{1,2} = 1.3$, $g_{SC} = 2$, $t = 0$, $J_{K_1} = 2.20$, $J_{K_2} = 3.23$, $\theta_B = 90°$ and $\phi_B = 90°$. The various arrows were added to indicate particular states for discussion in the text.

# Quantifying the quantum nature of high spin YSR excitations in transverse magnetic field

Niels P.E. van Mullekom[1], Benjamin Verlhac[1], Werner M.J. van Weerdenburg[1], Hermann Osterhage[1], Manuel Steinbrecher[1], Katharina J. Franke[2], A.A. Khajetoorians[1]*

**Affiliations:**

[1]*Institute for Molecules and Materials, Radboud University, Nijmegen, The Netherlands*

[2]*Fachbereich Physik, Freie Universität Berlin, Germany*
*corresponding author: <u>a.khajetoorians@science.ru.nl</u>

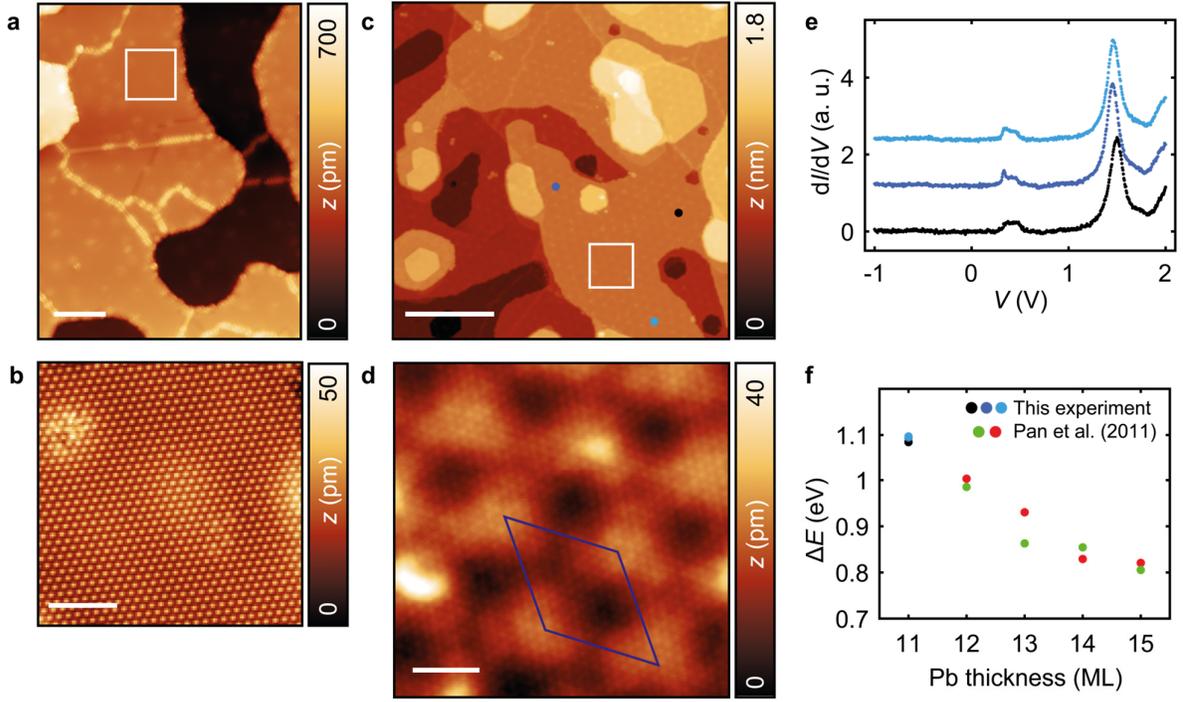

**Figure S1: Substrate characterization. (a)** Large-scale constant-current STM image of the reconstruction Si(111)-Ag($\sqrt{3} \times \sqrt{3}$) ($V_s = 1$ V, $I_t = 5$ pA, $T = 7$ K, scale bar: 20 nm). **(b)** Zoomed in constant-current STM image of the white square in **(a)** ($V_s = 1$ V, $I_t = 5$ pA, $T = 7$ K, scale bar: 5 nm). **(c)** constant-current STM image of a 11 ML Pb film on Si(111)-Ag($\sqrt{3} \times \sqrt{3}$) ($V_s = 90$ mV, $I_t = 5$ pA, $T = 30$ mK, scale bar: 20 nm). **(d)** constant-current STM image of the white square in **(c)** showing atomic resolution and the moiré-pattern. ($V_s = 3$ mV, $I_t = 6$ nA, $T = 30$ mK, scale bar: 2 nm). **(e)** STS measurements of the QWS at locations indicated in **(c)** by the colored dots, vertically offset for clarity ($V_s = 1$ V, $I_t = 10$ pA, $V_{mod} = 5$ mV, $T = 30$ mK). **(f)** Comparison of the difference in QWS energies in **(e)** with ref. 39 for Pb films on Si(111)-Ag($\sqrt{3} \times \sqrt{3}$) (red dots) and Pb films on Si(111) (green dots).

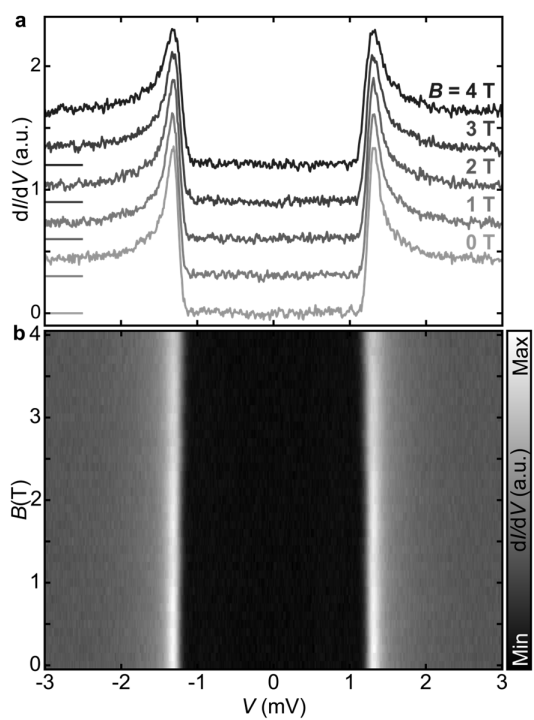

**Figure S2: SC gap spectra at variable magnetic field. (a)** Point spectra taken at transverse magnetic field strength indicated, vertically offset for clarity with zero indicated by horizontal lines. (**b**) a false-color plot of the STS spectra taken at $B_\parallel$ in steps of $\Delta B_\parallel = 0.1$ T, up to $B_\parallel = 4$ T. All data was measured with $V_S = 6$ mV, $I_t = 200$ pA, $V_{mod} = 20\ \mu$V, $T = 30$ mK and a W tip.

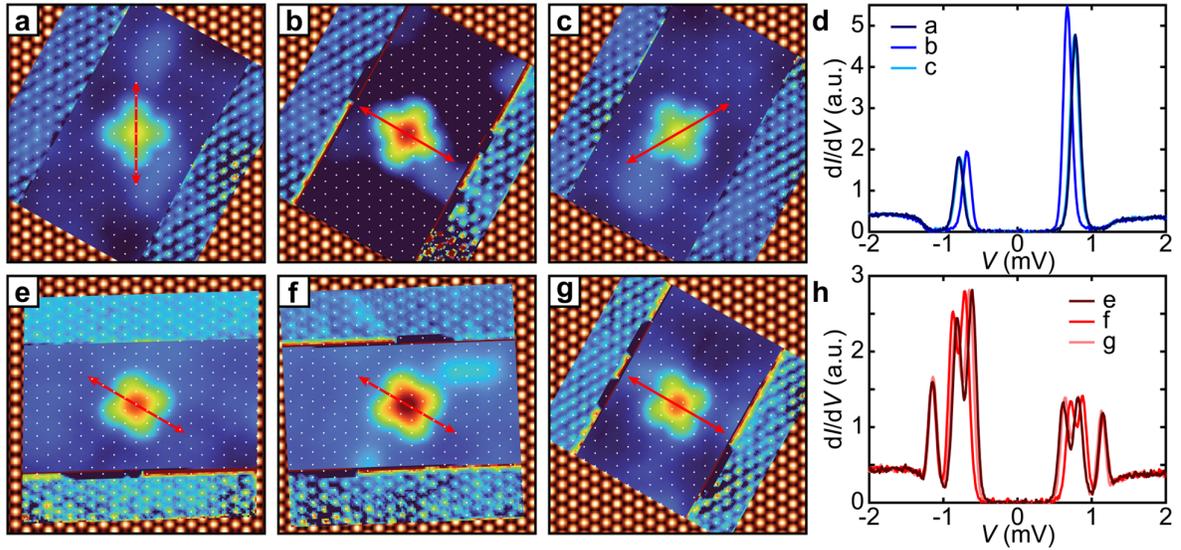

**Figure S3: Binding site analysis versus YSR spectrum. (a-c)** Processed constant-current STM-images of various MnPc1 on top of simulated Pb(111) lattice with atomic positions overlaid in white dots. The constant-current STM image was taken with two different parameter sets: for the atomic resolution away from the molecule ($V_S = 3$ mV, $I_t = 6$ nA) and with the molecule ($V_S = 90$ mV, $I_t = 20$ pA). The atomic resolution part was processed by separating it using a threshold, then flattened and scaled to enhance the contrast, and subsequently matched to the simulated lattice. Red arrows show ligand axis parallel to one of the high symmetry directions of Pb(111). **(d)** STS of MnPc shown in **(a-c)**, showing typical MnPc1 YSR states. **(e-g)** Same as **(a-c)**, but for MnPc2. The red arrows indicate the high symmetry direction of Pb(111) that bisects the ligand axes. **(h)** STS of MnPc shown in **(e-g)** showing typical MnPc2 YSR states. All spectra measured using $V_S = 6$ mV, $I_t = 200$ pA, $V_{mod} = 20 \, \mu$V.

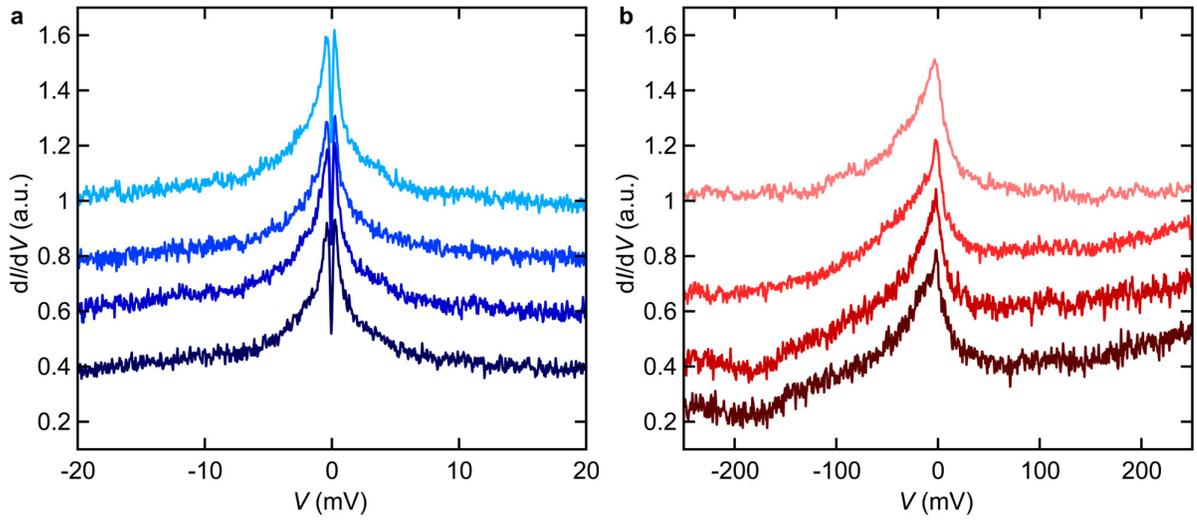

**Figure S4: Kondo spectra of MnPc.** The superconducting state of the Pb film is quenched by applying an out-of-plane magnetic field of $B_z = 0.5$ T. **(a)** STS of various MnPc1, showing a Zeeman-split Kondo resonance ($V_S = 20$ mV, $I_t = 200$ pA, $V_{mod} = 40$ $\mu$V). **(b)** STS of various MnPc2 featuring a broad Kondo resonance ($V_S = 250$ mV, $I_t = 100$ pA, $V_{mod} = 1$ mV).

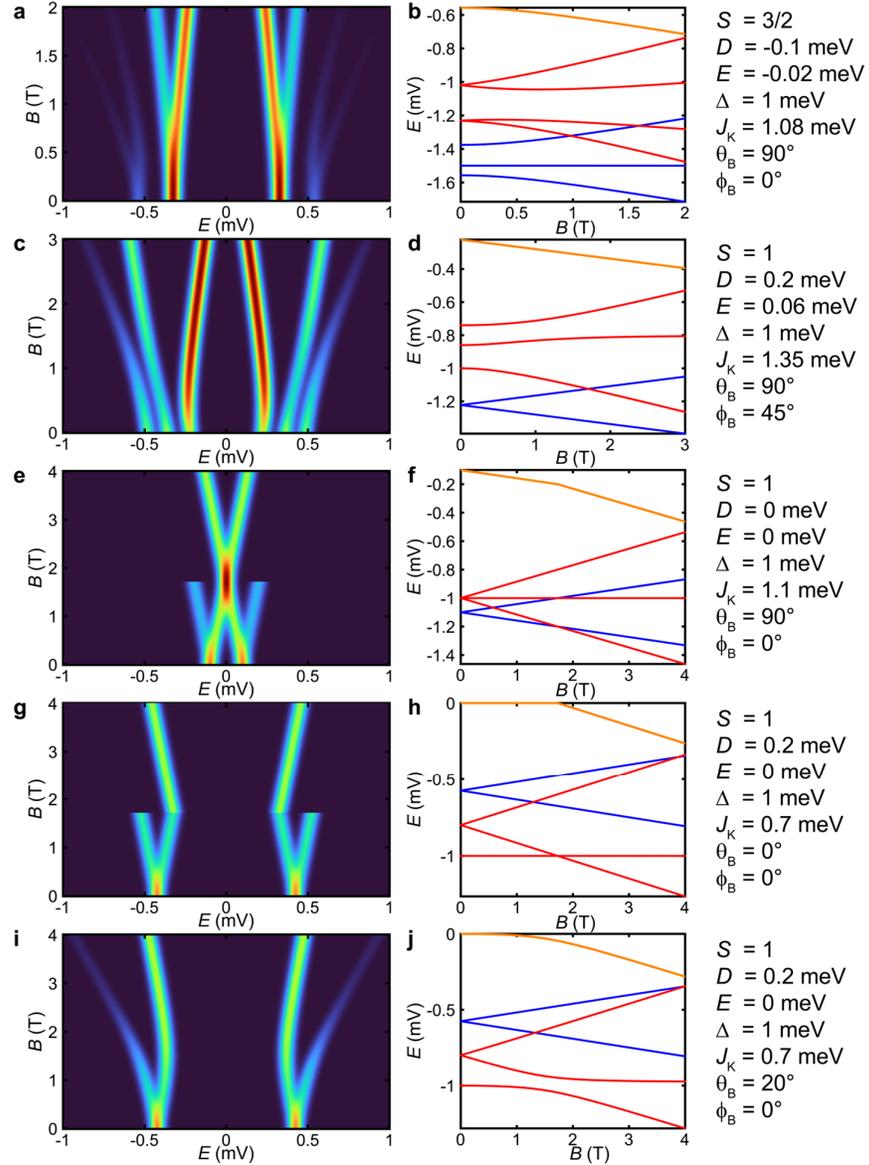

**Figure S5: Various calculations using the zero-bandwidth model, illustrating the various trends. (a-b)** An example of *B*-field dependent splitting of YSR excitations due to Kramer's degenerate excited states where the simulated spectra (left) and the energy level diagram (right) are illustrated. **(c-d)** An example of non-linear YSR excitations due to magnetic anisotropy where the simulated spectra (left) and the energy level diagram (right) are illustrated. **(e-f)** An example of a change in number of YSR excitations due to QPT in fermion parity where the simulated spectra (left) and the energy level diagram (right) are illustrated. **(g-h)** An example of a change in number of YSR excitations due to QPT in spin ground state where the simulated spectra (left) and the energy level diagram (right) are illustrated. **(i-j)** An example of broken spin rotation symmetry due to transverse magnetic field component where the simulated spectra (left) and the energy level diagram (right) are illustrated. In all energy level diagrams: red (blue) lines indicate states with $P = 1$ ($P = -1$), and orange lines correspond to $E_{gs} + \Delta$.

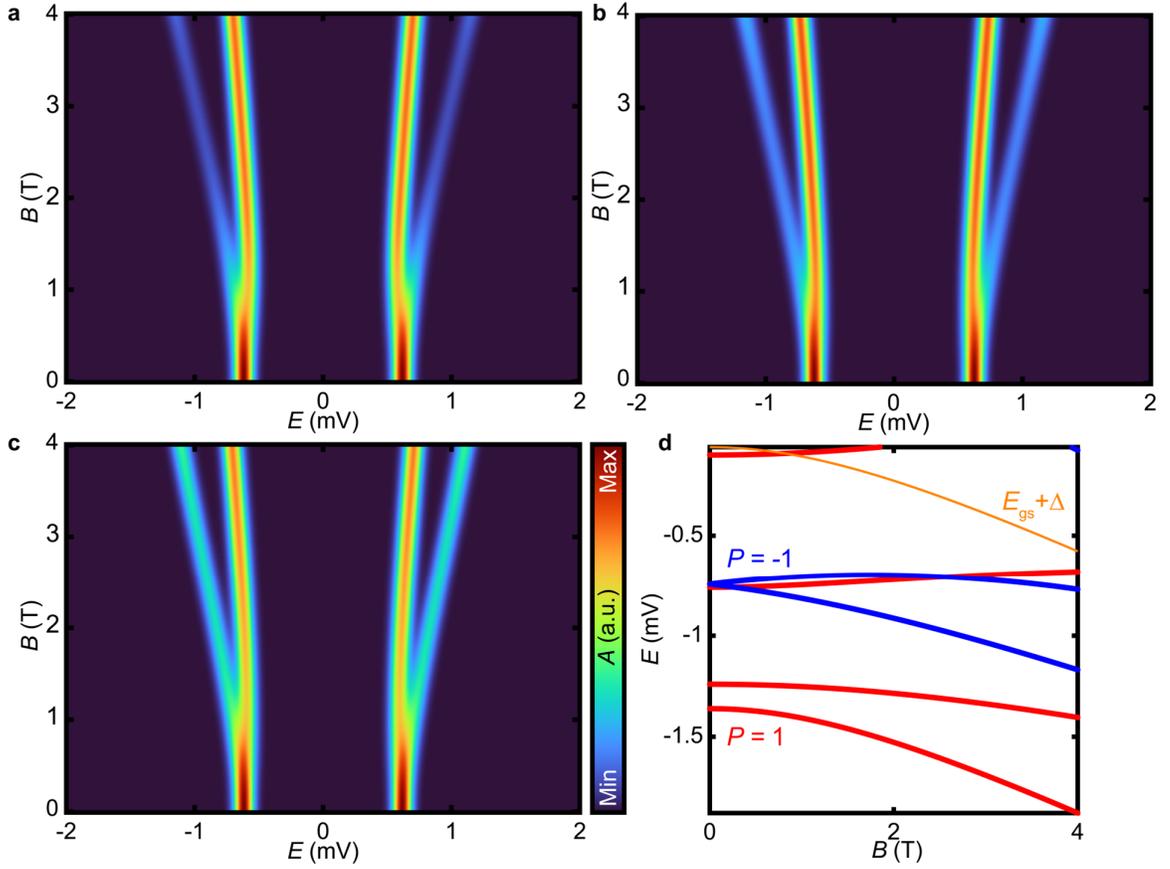

**Figure S6: Alternative YSR simulations for MnPc1. (a)** Reproduction of Fig. 3b plotted as reference. **(b)** Simulation with easy-plane anisotropy instead, using model parameters: $S = 1$, $g = 2$, $D = 0.3$ mV, $E = 0$ meV, $\Delta = 1.3$ mV, $J_K = 0.86$ mV, $\theta_B = 90°$, $\phi_B = 0°$. **(c)** Simulation with different total spin instead, using model parameters: $S = 2$, $g = 2$, $D = 0.3$ mV, $E = 0.08$ meV, $\Delta = 1.3$ mV, $g_{SC} = 2$, $J_K = 0.56$ mV, $\theta_B = 90°$ and $\phi_B = 0°$. **(d)** Partial energy level diagram of **(c)**, showing lowest energy states up to the ground state, with the orange line being $E_{gs} + \Delta$.

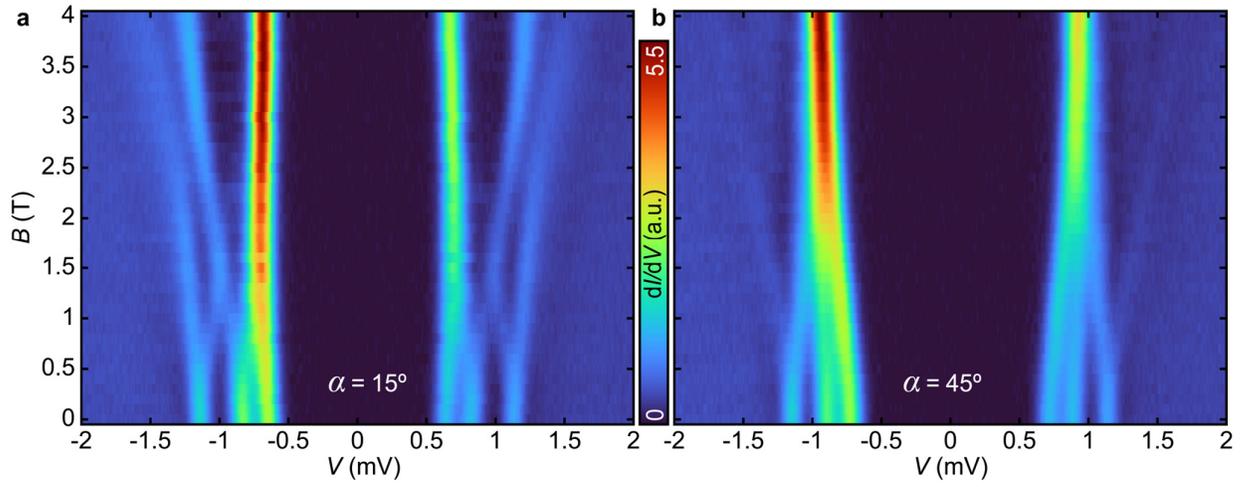

**Figure S7: Transverse magnetic field dependence of the YSR excitations of MnPc2. (a)** A false-color plot of the STS spectra of MnPc2($\alpha = 15°$) taken at $B_\parallel$ in steps of $\Delta B_\parallel$ =0.1 T , up to $B_\parallel = 4.0$ T ($V_S = 6$ mV, $I_t = 200$ pA, $V_{mod} = 20$ $\mu$V). **(b)** A false-color plot of the STS spectra of MnPc2($\alpha = 45°$) taken at $B_\parallel$ steps of $\Delta B_\parallel = 0.1$ T, up to $B_\parallel = 4.0$ T.

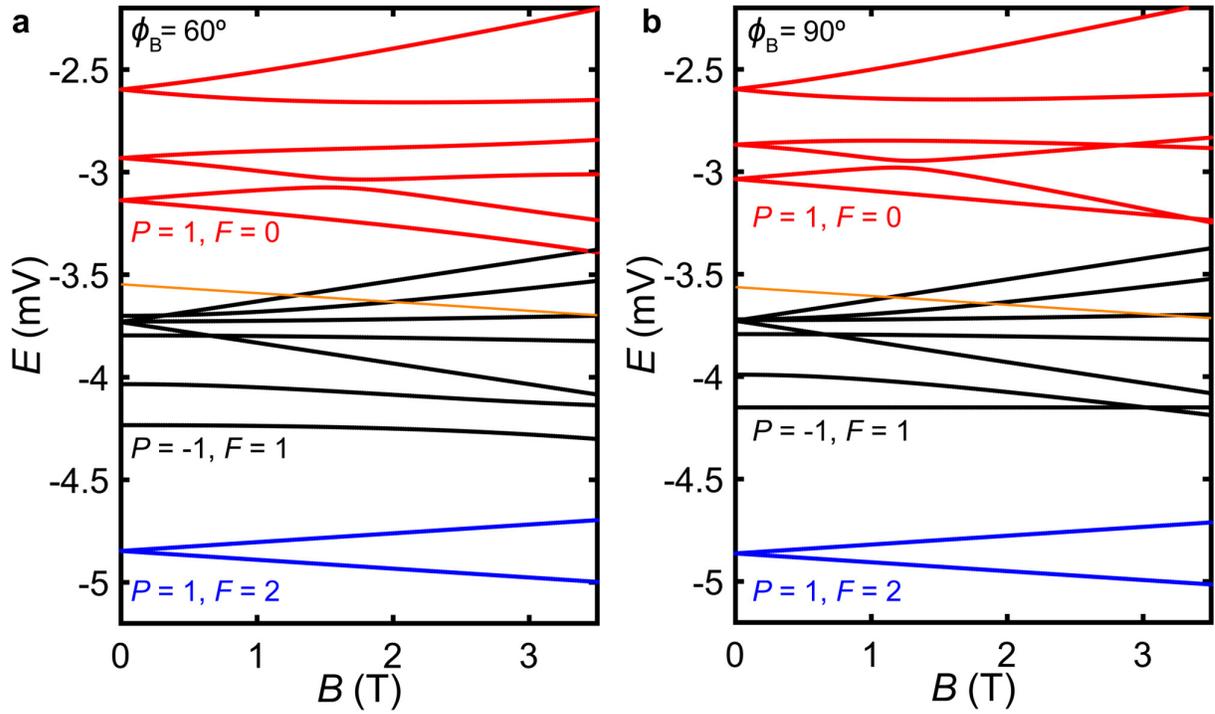

**Figure S8: Energy level diagram of MnPc2 YSR simulations in Fig. 4. (a-b)** The energy level diagram corresponding to Fig. 4c. and Fig. 4d, respectively. The orange line corresponds to the ground state energy $E_{gs} + \Delta$. $P$ is the fermion parity and $F$ is the number of bound quasiparticles.